\documentclass[journal,onecolumn]{IEEEtran}
\usepackage{amsthm}
\usepackage{subfigure}
\usepackage{graphicx}
\usepackage{amssymb,amstext,amsmath}
\usepackage{float}
\usepackage{bbm}
\usepackage{bm}
\usepackage{multirow}
\usepackage{makecell}
\usepackage{color}
\usepackage{caption}
\usepackage{verbatim}
\usepackage{float}
\usepackage{mathtools}
\usepackage{amssymb}
\usepackage{algorithm}
\usepackage{color}
\usepackage{booktabs}
\usepackage{tikz}
\usepackage{balance}
\usetikzlibrary{arrows}

\newcommand{\x}{{\bf x}}

\newcommand{\bSigma}{\boldsymbol\Sigma}
\newcommand{\bmu}{\boldsymbol\mu}

\def\Item$#1${\item $\displaystyle#1$
\hfill\refstepcounter{equation}(\theequation)}

\newcommand\blfootnote[1]{%
  \begingroup
  \renewcommand\thefootnote{}\footnote{#1}%
  \addtocounter{footnote}{-1}%
  \endgroup
}

\usepackage[noend]{algpseudocode}

\usepackage[skip=2pt]{caption}

\makeatletter
\def\BState{\State\hskip-\ALG@thistlm}
\makeatother

\makeatletter
\newcommand{\skipitems}[1]{%
	\addtocounter{\@enumctr}{#1}%
}
\makeatother

\begin{document}

\title{Robust Covariance Adaptation in Adaptive Importance Sampling} 
\author{Yousef El-Laham,*  V\'{i}ctor Elvira,$^\dagger$ and M\'{o}nica F. Bugallo* \\
* Dept. of Electrical and Computer Engineering, Stony Brook University, New York (USA) \\
$^\dagger$ IMT Lille Douai \& CRIStAL (UMR CNRS 9189), Villeneuve d'Ascq, France.}
\markboth{}%
{Shell \MakeLowercase{\textit{et al.}}: Bare Demo of IEEEtran.cls for Journals}
% insert page header and footer here for IEEE PDF Compliant

\maketitle

% FOOTNOTE 
\blfootnote{M. F. B. thanks the support of the National Science Foundation under Award CCF-1617986. V. E. acknowledges support from the ANR under PISCES project (ANR-17-CE40-0031-01), and from the French-American Fulbright Commission under the Fulbright scholarship. \copyright2018
	IEEE}

% ABSTRACT
\begin{abstract}
Importance sampling (IS) is a Monte Carlo methodology that allows for approximation of a target distribution using weighted samples generated from another proposal distribution. Adaptive importance sampling (AIS) implements an iterative version of IS which adapts the parameters of the proposal distribution in order to improve estimation of the target. While the adaptation of the location (mean) of the proposals has been largely studied, an important challenge of AIS relates to the difficulty of adapting the  scale parameter (covariance matrix).  In the case of weight degeneracy, adapting the covariance matrix using the  empirical covariance results in a singular matrix, which leads to poor performance in subsequent iterations of the algorithm. In this paper, we propose a novel scheme which exploits recent advances in the IS literature to prevent the so-called weight degeneracy. The method efficiently adapts the covariance matrix of a population of proposal distributions and achieves a significant performance improvement in high-dimensional scenarios. We validate the new method through computer simulations.
\end{abstract}

% KEYWORDS

% INTRODUCTION 
\section{Introduction}
\label{sec:intro}

Importance sampling (IS) is a powerful Monte Carlo tool that allows  to learn properties of a distribution (called target) by utilizing weighted samples drawn from another distribution (called proposal) \cite{Robert2004}. The choice of the proposal distribution in IS algorithms is critical for successful performance. In particular, one issue that can arise in IS when a poorly chosen proposal is used  is \textit{weight degeneracy}, which relates to the problem of only a few samples having significant weights that contribute to approximation of the target distribution. High mismatch between the proposal distribution and the target distribution yields a high variance IS estimator. Alternatively, one can use a method known as multiple importance sampling (MIS), in which samples are drawn from multiple proposal distributions \cite{Veach95,Owen2000}. MIS methods have an advantage over static IS methods as
%in that 
%the generated samples come from multiple distributions 
they show larger sample diversity and alternative weighting schemes can be applied to reduce the variance of the IS estimator \cite{elvira2015generalized,elvira2015efficient,elvira2016heretical}.
%as proved in \cite{elvira2015generalized} and developed in \cite{elvira2015efficient,elvira2016heretical} }. 
Adaptive importance sampling (AIS) methods were introduced in order to adapt the proposal distribution through an iterative learning algorithm \cite{Oh1992}. While AIS methods have shown success in many practical applications \cite{Au1999,Cappe2004}, major challenges of the method become apparent when considering multimodal and highly nonlinear target distributions. Moreover, AIS also suffers from the curse of dimensionality \cite{Bengtsson2008}, whereby the issue of weight degeneracy intensifies and leads to reduced sample diversity in subsequent iterations of the algorithm. 

Some advances have been made to AIS schemes to increase the diversity of the generated samples and limit the effects of weight degeneracy. The use of deterministic mixture weights \cite{Owen2000} in the population Monte Carlo scheme (DM-PMC) \cite{Elvira2017} significantly reduces the variance of the IS estimator and maintains the diversity of the proposals through each iteration. Unfortunately, the DM-PMC algorithm does not adapt the covariance matrices of the proposal distributions. Alternatively, the nonlinear population Monte Carlo (N-PMC) scheme applies a nonlinear transformation of the importance weights, similar to the approaches of \cite{ionides2008truncated,Koblents2013,vehtari2015pareto}.  The weight transformation guarantees representation of more samples for proposal adaptation in the case of weight degeneracy. However, in high dimensional spaces, the effect of the weight transformation may lead to minuscule adaptation of the proposal.  Other AIS methods such as mixture population Monte Carlo (M-PMC) \cite{Cappe2008}, adaptive multiple importance sampling (AMIS) \cite{cornuet2012adaptive}, and adaptive population importance sampling (APIS) \cite{Martino2015}, have also been proposed with some variations. Some of these AIS methods also adapt the covariance matrix, but the stability of the update is not guaranteed. In this paper, we propose a novel and robust method for adapting a set of proposal distributions, called \textit{covariance adaptive importance sampling} (CAIS), which tackles the inherent problems posed by high-dimensional targets. Specifically, we condition the covariance adaptation of each proposal on a local measurement of the effective sample size in order to achieve robustness. 

The rest of the paper is organized as follows. In Section \ref{sec:problem formulation}, we describe the problem. Section \ref{sec:novelty} describes the proposed novel approach for adapting the proposal distributions. We show simulation results in Section \ref{sec:simulations} and conclude the paper in Section \ref{sec:conclusion}. 

\vspace*{-.3cm}
% PROBLEM FORMULATION
\section{Problem Formulation}
\label{sec:problem formulation}

We introduce the estimation problem in a Bayesian context, with the goal of computing the posterior distribution $p(\textbf{x}|\textbf{y})$,  where $\textbf{x}\in\mathbb{R}^{d_x}$ is a $d_x$-dimensional vector of unknown parameters related to a set of available observations  $\textbf{y}\in\mathbb{R}^{d_y}$.  It follows from Bayes' rule that,
\begin{equation}
	p(\textbf{x}|\textbf{y}) \propto p(\textbf{y}|\textbf{x})p(\textbf{x})\equiv\pi(\textbf{x}),
\end{equation}
where $p(\textbf{y}|\textbf{x})$ is the likelihood function of the data, while $p(\textbf{x})$ is a prior distribution of the vector of unknown parameters. The goal is in computing the expectation of some function $g(\textbf{X})$ w.r.t. $p(\textbf{x}|\textbf{y})$, such that
\begin{equation}
	I=\mathbb{E}(g(\textbf{X}))=\int_{-\infty}^{\infty} g(\textbf{x}) p(\textbf{x}|\textbf{y}) d \textbf{x}.
\end{equation}
In particular, we may be interested in computing quantities such as the normalizing constant $Z=\int \pi(\textbf{x})d\textbf{x}$ or the mean of the posterior distribution.  To that end, basic implementations of IS approximate $p(\textbf{x}|\textbf{y})$ using samples drawn from a proposal distribution $q(\textbf{x})$ that are weighted properly.

\section{Novel Covariance Adaptation Strategy}
\label{sec:novelty}
\begin{table}[t]
	\centering
	\caption{A basic framework for AIS methods.}
	\label{tab:basic_ais}
	%\resizebox{0.9\columnwidth}{!}{
	\begin{tabular}{l}
		\midrule
		\textbf{1. Initialization:} Select the initial proposal $q(\textbf{x}|\boldsymbol{\mu}_1,\boldsymbol{\Sigma}_1)$ \\
		\textbf{2. For $i=1,...,I$}
		\\
		\quad a. Draw $M$ samples from the proposal,
		\\
		\quad\quad\quad\quad\parbox{2cm}{
			\begin{equation*}
			\textbf{x}_i^{(m)}\sim q(\textbf{x};\boldsymbol{\mu}_{i},\boldsymbol{\Sigma}_{i}), \quad\quad m=1,...,M.
			\end{equation*}
		}
		\\
		\quad b. Compute the standard importance weights,
		\\
		\quad\quad\quad\quad\parbox{2cm}{
			\begin{equation*}
			w_i^{(m)}=\frac{\pi(\textbf{x}_i^{(m)})}{q(\textbf{x}_i^{(m)};\boldsymbol{\mu}_{i},\boldsymbol{\Sigma}_{i})}, \quad m=1,...,M,
			\end{equation*}}
		\\
		\quad\quad \ and normalize them, $\bar{w}_i^{(m)}=\frac{w_i^{(m)}}{\sum_{j=1}^M w_i^{(j)}},$ with \\\quad\quad \ $m=1,...,M.$
		\\
		\quad c. Adapt the proposal parameters $\boldsymbol{\mu}_{i+1}$ and $\boldsymbol{\Sigma}_{i+1}$. \\ 
		\textbf{3.} Return the pairs $\{\textbf{x}_i^{(m)},w_i^{(m)}\}$ for $m=1,...,M$ and \\ \quad \ $i=1,...,I$.
		\\
		\bottomrule
	\end{tabular}
	% }
	
\end{table}
We assume a family of proposal distributions, $q(\textbf{x};\boldsymbol{\mu},\boldsymbol{\Sigma})$, where the parameters $\boldsymbol{\mu}$ and $\boldsymbol{\Sigma}$ are the mean and covariance matrix, respectively.  Table \ref{tab:basic_ais} summarizes a basic AIS framework, where the proposal is adapted over $I$ iterations.
% General Formulation
\subsection{General Formulation}
\label{subsec: general_formulation}
For a multivariate normal distribution, the maximum likelihood (ML) estimate of the covariance matrix for $M$ unweighted samples with mean $\hat\bmu_{ML}$ is given by,
\begin{equation}
\label{eq: ML_estimate_cov}
\hat\bSigma_{ML}=\frac{1}{M}\sum_{m=1}^M (\x^{(m)}-\hat\bmu_{ML})(\x^{(m)}-\hat\bmu_{ML})^\mathsf{T}.
\end{equation}
It is well-known that if $\x\in\mathbb{R}^{d_x}$ and $M < d_x$, then $\text{rank}(\hat\bSigma_{ML})<d_x$ and $\hat\bSigma_{ML}$ cannot be inverted \cite{Ledoit2004}. Analogous to the case of unweighted samples, suppose that we are given a set of $M$ weighted samples $\{\x^{(m)},\bar{w}^{(m)}\}_{m=1}^M$. If the weighted mean of the samples is $\hat\bmu_W$, then the weighted empirical covariance is given by,
\begin{equation}
\label{eq: weighted_empirical_cov}
\hat\bSigma_W=\sum_{m=1}^M \bar{w}^{(m)}(\x^{(m)}-\hat\bmu_{W})(\x^{(m)}-\hat\bmu_{W})^\mathsf{T}.
\end{equation}
Consider the set of indices $K=\{k \ | \ \bar{w}^{(k)}<\epsilon\}$. As $\epsilon\rightarrow 0$, we have that $\hat\bSigma_W\approx\sum_{k\notin K} \bar{w}^{(k)}(\x^{(k)}-\hat\bmu_{W})(\x^{(k)}-\hat\bmu_{W})^\mathsf{T}$ and $\text{rank}(\hat\bSigma_W)\leq |K^C|$. If $|K^C|<d_x$, then \eqref{eq: weighted_empirical_cov} yields a singular matrix. For this reason, adaptation of the covariance matrix in AIS is challenging due to the chance of  weight degeneracy of the samples at each iteration. 

We can characterize the degeneracy of the importance weights in AIS at any iteration through the effective sample size (ESS) \cite{kong92}, which can approximated as,

\begin{equation}
\label{ESS_exact}
\hat \eta_i=\frac{1}{\sum_{m=1}^M(\bar{w}_i^{(m)})^2}.
\end{equation}

It is important to note that the approximation in \eqref{ESS_exact} is only true under certain conditions \cite{Martino2017}. Moreover, other approximations for ESS can be used, such as $\tilde\eta_i=\max(\bar w_i^{(m)})^{-1}$, though for our experiments we use eq. \eqref{ESS_exact}. Under the assumption that $\text{rank}(\bSigma_{ML})=M$ at each AIS iteration (i.e. large enough variability in the generated samples), our criterion for robust covariance adaptation rests upon guaranteeing that the ESS at each iteration of the algorithm satisfies some lower bound condition. 

%Essentially, in the cases of extreme degeneracy, we relax the adaptation of the covariance matrix by distributing the weights more evenly among the samples. This conditioning on the ESS guarantees that we only use the standard importance weights for covariance adaptation when the proposal distribution is close enough to the target distribution.

% Table with the covariance adapting scheme

% Presentation of algorithm
\subsection{Algorithm Description}
In this section, we formalize a novel algorithm, \textit{covariance adaptive importance sampling} (CAIS), which guarantees robust covariance adaptation for a population of proposal distributions. 
\begin{enumerate}

% STEP 1: INITIALIZATION
\item \textbf{Initialization:} Initialize the proposal parameters $\bmu_{1,d}, \bSigma_{1,d}$ for $d=1,...,D$ and set $i=1$. 

% STEP 2: DRAW NEW SET OF SAMPLES
\item \textbf{Generate samples:} Draw $N$ samples from each mixand,
\begin{equation*}
\x_{i,d}^{(n)}\sim q(\x; \bmu_{i,d}, \bSigma_{i,d}), \quad n=1,...,N, \quad d=1,...,D.
\end{equation*}

% STEP 3: COMPUTE UNNORMALIZED WEIGHTS
\item \textbf{Compute weights:} Evaluate the standard importance weights of each sample,    
\begin{equation*}
w_{i,d}^{(n)}=\frac{\pi(\x_{i,d}^{(n)})}{q(\x_{i,d}^{(n)}; \bmu_{i,d}, \bSigma_{i,d})}, \ \ n=1,...,N, \ \ d=1,...,D,
\end{equation*}
and normalize locally,
\begin{equation}
\label{eq: weight_normalization}
\bar{w}_{i,d}^{(n)}=\frac{w_{i,d}^{(n)}}{\sum_{j=1}^{N}w_{i,d}^{(j)}}, \ \  n=1,...,N, \ \ d=1,...,D.
\end{equation}
We note that alternative weighting schemes can be used for target approximation, such as the deterministic mixture (DM) weights presented in \cite{Owen2000,Veach95,elvira2015generalized}. 
 
%STEP 4: Compute local ESS for each set of samples
\item \textbf{Compute local ESS:} For each set of samples and normalized weights, $\{\x_{i,d}^{(n)},\bar w_{i,d}^{(n)}\}_{n=1}^N$, compute the local ESS, $\hat \eta_{i,d}$, by applying eq. \eqref{ESS_exact}.

% REVISED VERSION
% STEP 5: Update the mean parameter
\item \textbf{Update mean:}
Update the mean of each mixand by taking the weighted sample mean,
\begin{equation}
\label{eq: weighted_mean}
\bmu_{i+1,d}=\sum_{n=1}^{N}\x_{i,d}^{(n)}\bar{w}_{i,d}^{(n)}, \quad d=1,...,D.
\end{equation}

\item \textbf{Update covariance matrix:} 
For $d=1,...,D$, 
\begin{enumerate}
	\item If $\hat\eta_{i,d}\geq N_{T}$, update the covariance matrix as, 
	\begin{equation*}
	\label{eq: weighted_empirical_cov_algorithm_untransformed}
	\bSigma_{i+1,d}=\sum_{n=1}^N \bar{w}_{i,d}^{(n)}(\x_{i,d}^{(n)}-\bmu_{i,d})(\x_{i,d}^{(n)}-\bmu_{i,d})^\mathsf{T}.
	\end{equation*}
    \item If $\hat\eta_{i,d}< N_{T}$, transform the importance weights as $w_{i,d}^{(n)*}=\psi(w_{i,d}^{(n)})$ for $n=1,...,N$. Normalize the transformed weights by applying eq. \eqref{eq: weight_normalization}. Compute the tempered mean $\bmu_{i,d}^*$ by applying eq. \eqref{eq: weighted_mean} given the normalized transformed weights. Update the covariance matrix as, 
\begin{equation*}
\label{eq: weighted_empirical_cov_algorithm_transformed}
\bSigma_{i+1,d}=\sum_{n=1}^N \bar{w}_{i,d}^{(n)*}(\x_{i,d}^{(n)}-\bmu_{i,d}^*)(\x_{i,d}^{(n)}-\bmu_{i,d}^*)^\mathsf{T}.
\end{equation*}
	\end{enumerate}

% % STEP 5: ADAPT PARAMETERS 
% \item \textbf{Adapt proposal parameters:} Update the mean of each mixand by taking the weighted sample mean,
% \begin{equation*}
% \label{eq: weighted_mean}
% \bmu_{i+1,d}=\sum_{n=1}^{N}\x_{i,d}^{(m)}\bar{w}_{i,d}^{(m)}, \quad d=1,...,D.
% \end{equation*}
% For $d=1,...,D$, if $\eta_{i,d}\geq N_{T}$, set $\bar{w}_{i,d}^{(n)*}=\bar{w}_{i,d}^{(n)}$ for $n=1,..,N$ and  $\bmu_{i,d}^*=\bmu_{i+1,d}$. Otherwise, compute the normalized transformed weights $\bar{w}_{i,d}^{(n)*}=\frac{\psi({w}_{i,d}^{(n)})}{\sum_{j=1}^N\psi({w}_{i,d}^{(j)})}$ for $n=1,...,N$ and $\bmu_{i,d}^*$ by applying equation \eqref{eq: weighted_mean} given the normalized transformed weights.  The covariance matrix is adapted as, 
% \begin{equation}
% \label{eq: weighted_empirical_cov_algorithm}
% \bSigma_{i+1,d}=\sum_{n=1}^N \bar{w}_{i,d}^{(n)*}(\x_{i,d}^{(n)}-\bmu_{i,d}^*)(\x_{i,d}^{(n)}-\bmu_{i,d}^*)^\mathsf{T},
% \end{equation}
% for $d=1,...,D$. 

% CHECK STOPPING CONDITION
\item \textbf{Check stopping condition:} If $i=I$, then return $\{\x_{i,d}^{(n)},w_{i,d}^{(n)}\}$ for $d=1,...,D$ and $n=1,...,N$. Otherwise, set $i=i+1$ and go to Step 2.      
\end{enumerate}

\vspace*{-.3cm}
%% REVISED SECTION - ADD ALGORITHM SUMMARY

\subsection{Algorithm Summary}
\label{subsec: algorithm_summary}
The novel technique adapts both the mean and covariance matrix for a population of proposal distributions. We emphasize that the untransformed weights are used in order to adapt the mean of each proposal, while the covariance adaptation is conditioned on the local ESS of the samples drawn by each proposal. If the local ESS is above a threshold $N_T$, the untransformed weights are used for the covariance matrix adaptation. Otherwise, the importance weights are transformed using a weight transformation function $\psi(\cdot)$ and the covariance is adapted using the transformed weights, yielding a wider and hence more stable proposal. 
% Choice of Weight Transformation Function
\subsection{Choice of the Weight Transformation Function}
\label{subsec: choice_of_weight_transformation}
% Weight Clipping  
\subsubsection{Weight Clipping}
\label{subsubsec: weight_clipping}
One choice for $\psi(\cdot)$ is the clipping function utilized in N-PMC.  N-PMC transforms the weights by building a sorted permutation of the unnormalized weights $\{w^{(n)}\}_{n=1}^N$ as
	\begin{equation}
    \label{eq: permute}
    w^{(k_1)}\geq w^{(k_2)}\geq...\geq w^{(k_N)}, 
    \end{equation}
and then transforming the subset of the greatest $1<N_{T}<N$ weights as 
	\begin{equation}
   	\label{eq: clipping} 
   	\psi(w^{(n)})=\textrm{min}\big(w^{(n)},w^{(k_{N_T})}\big). 
	\end{equation}
This choice of $\psi(\cdot)$ guarantees that the ESS determined by the transformed weights is bounded below by $N_T$. Intuitively, in the case of extreme degeneracy, this can be interpreted as adapting the covariance matrix using the ML estimate of the covariance considering only the $N_T$ highest weighted samples.

% Power Function
\subsubsection{Weight Tempering}
\label{subsubsec: power_function}

While \eqref{eq: clipping} guarantees that the minimum ESS condition is satisfied, it also forces the $N_T$ largest weights to take identical value, eliminating any information about which of these weights are most representative of the target distribution. An alternative choice, which offers more flexibility, is the weight tempering function,
\begin{equation}
\label{eq: power_function}
\psi(w_i^{(n)})=\big(w_i^{(n)}\big)^{\frac{1}{\gamma}}, \quad \gamma \geq 1, 
\end{equation}
where $\gamma$ controls the transformation of the importance weights.

Note that all strategies transform the importance weights but preserve the position of the samples. More advanced mechanisms could be devised by implementing covariance estimators where the samples are also transformed, following the ideas of \cite{martino2017group}.
\subsection{Choice of Parameters}
\label{subsec: choice_Of_threshold}
The choice of the parameter $N_T$ depends on the dimension of the target, $d_x$. In order to guarantee stability of the covariance update, we must have that $N_T>d_x$.  However, we note that under poor initialization of the proposal distribution, a choice of large $N_T$ (relative to $N$)  will force a strong nonlinear transformation of the importance weights, regardless of the $\psi(\cdot)$ chosen. Consequently, the adapted covariance matrix will not differ from the one from the previous iteration, resulting in slower convergence to the true target. 

The weight tempering function should be such that the ESS of the transformed weights is $\hat\eta^*\approx N_T$. In order to find the optimal $\gamma$, it is sufficient to employ a combination of loose grid search and fine grid search, similar to the approach used in \cite{Chih-WeiHsuChih-ChungChang2008}. For $\gamma$ that is $\epsilon$-suboptimal (i.e. $|N_T-\hat\eta^*|<\epsilon$), the computational complexity of a standard one-dimensional grid-search is $\mathcal{O}(\frac{1}{\epsilon})$. For a reasonable choice of $\epsilon$ (not too small),  this is overshadowed by the computational complexity of each covariance matrix update, which is $\mathcal{O}(Nd_x^2)$. \\

% Numerical Examples
\section{Numerical Examples}
\label{sec:simulations}

% Example 1: Unimodal Target using poorly initialized proposal distribution
\subsection{Example 1: Unimodal Gaussian Target}
\label{subsec: example_1}
% Big figure for example 1   
\begin{figure*}[t]
	\centering
	\begin{minipage}[b]{0.3\linewidth}
		\includegraphics[width=5cm]{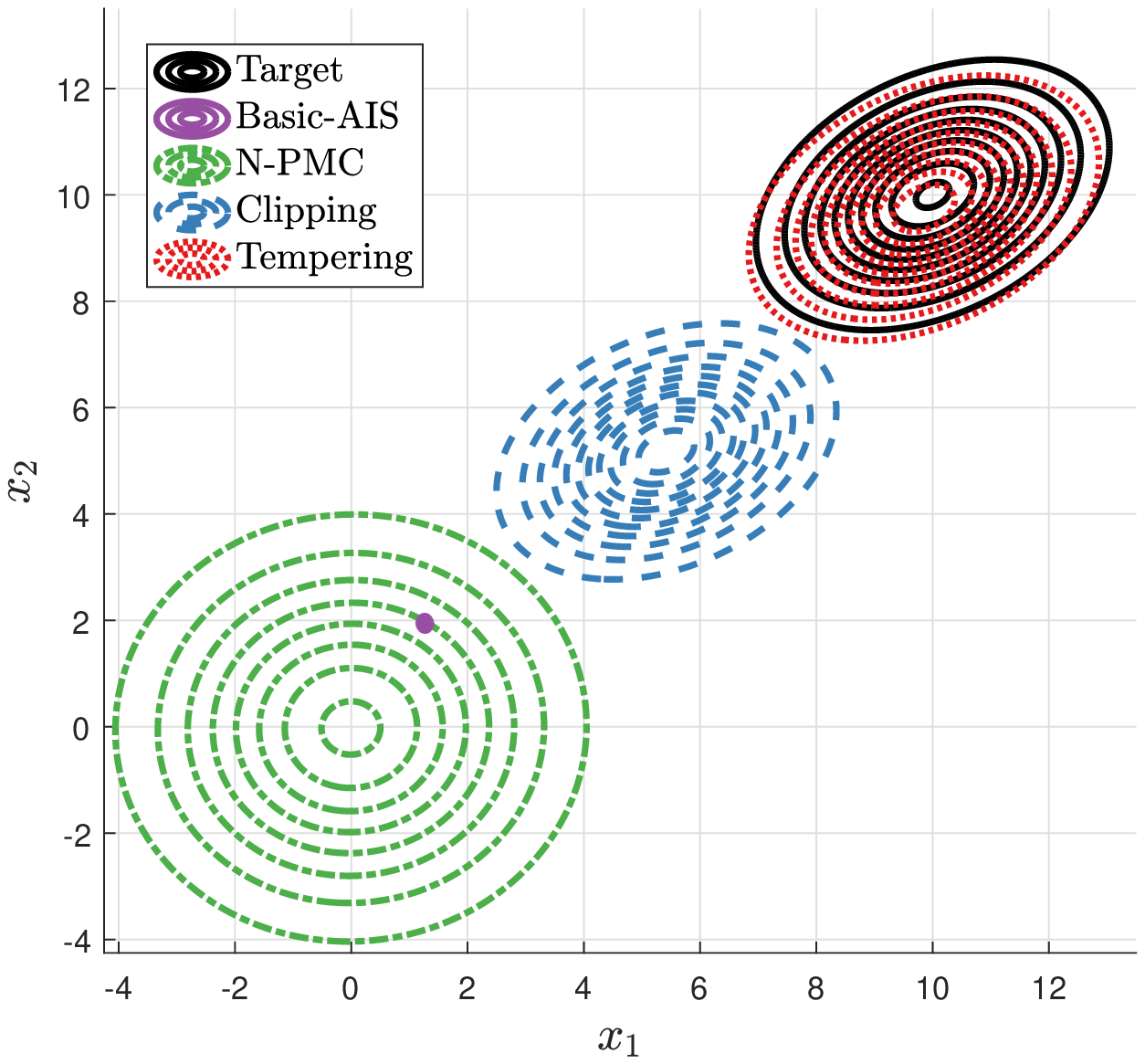}
	\end{minipage}
	\begin{minipage}[b]{0.3\linewidth}
		\includegraphics[width=5cm]{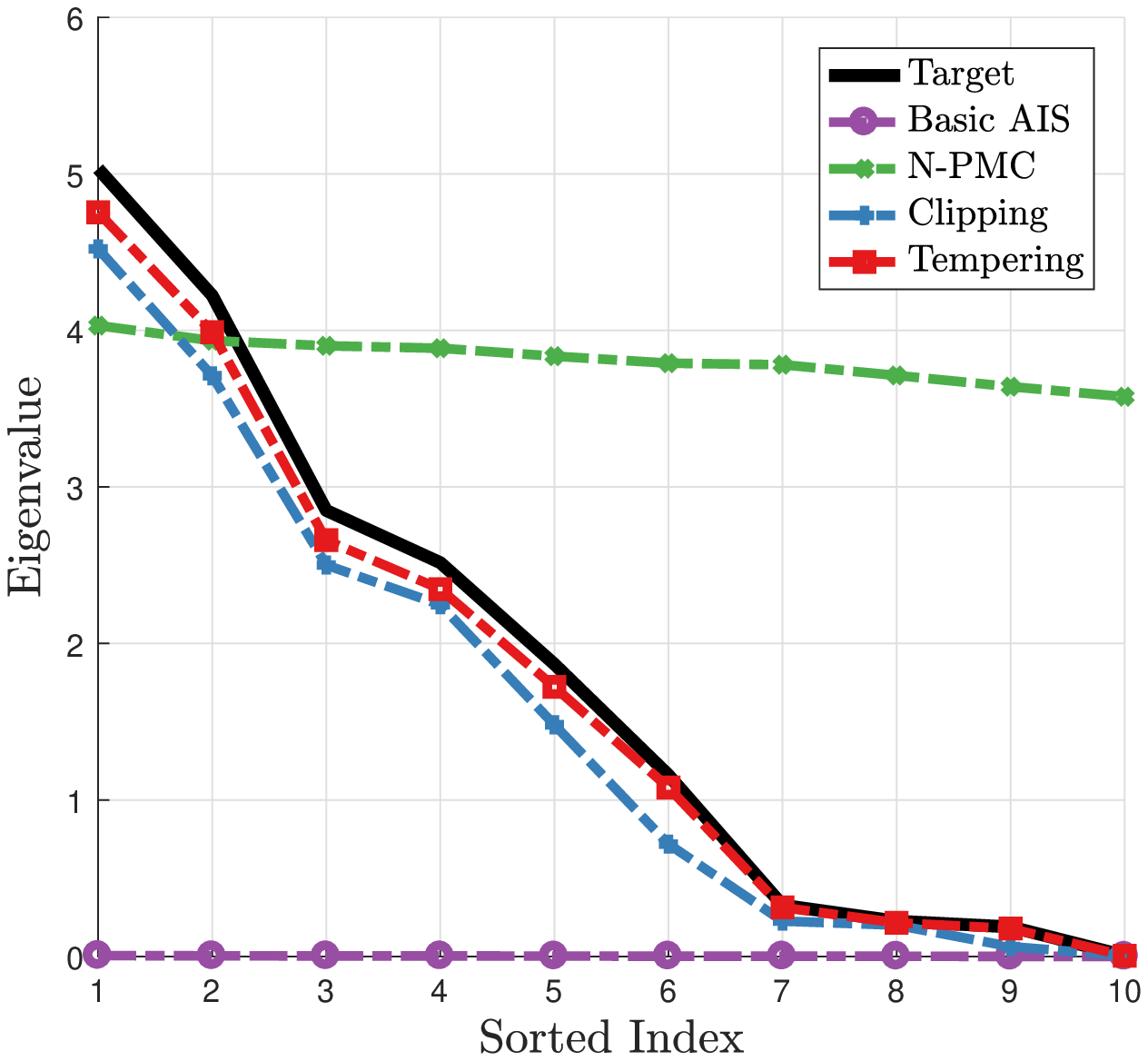}
	\end{minipage}
	\begin{minipage}[b]{0.3\linewidth}
		\includegraphics[width=5cm]{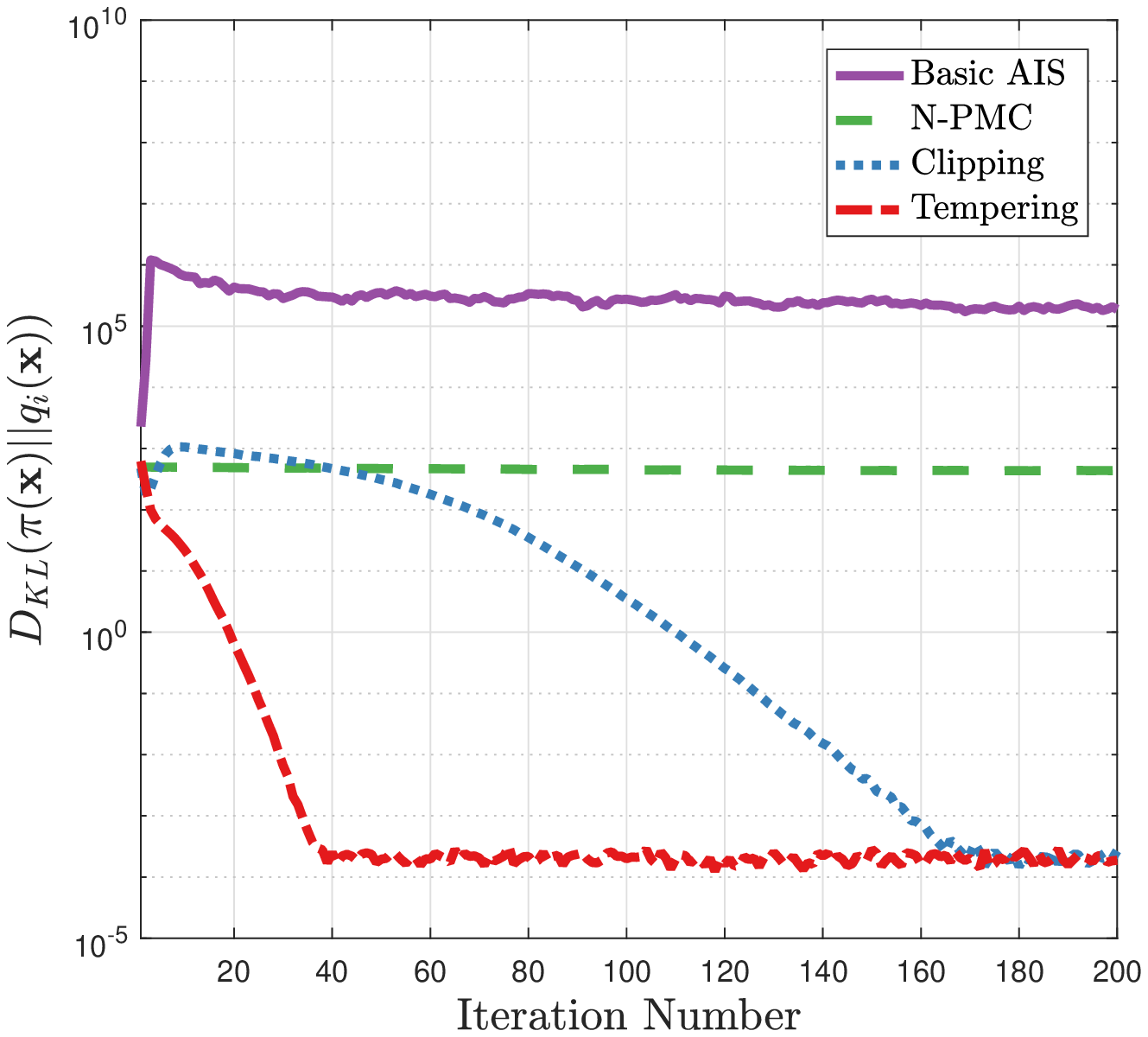}
	\end{minipage}
	\caption{(Example 1) The left figure shows the slice $[x_1, x_2]^\top$ of the adapted proposal distribution after $25$ iterations. Note that the basic AIS provides a poor covariance adaptation with the proposal degenerating to a delta in a part of space which is far away from the target distribution. The middle figure compares the sorted eigenvalues of the target distribution's covariance matrix with that of the proposal distribution after $25$ iterations. The right figure shows the evolution of the Kullback-Liebler divergence between the target distribution and the proposal distribution over $200$ iterations. }\label{fig: example_1_stuff}
\end{figure*}

We consider a toy example to demonstrate fundamental differences between existing AIS schemes which adapt the covariance matrix and the novel strategy. In particular, the target of interest is a 10-dimensional unimodal Gaussian distribution with $\nu_i=10$ for $i=1,...,10$ and a non-diagonal covariance matrix randomly generated from a Wishart distribution. We use a single proposal distribution for each method  and initialize  the mean as $\bmu_{1}=\boldsymbol{0}$ and the covariance matrix as $\boldsymbol{\Sigma}_{1}=4\times\mathbb{I}_{10}$, where $\mathbb{I}_{10}$ is the $10\times10$ identity matrix. This is a poor initialization since it is far from the mean of the target distribution. We draw $N=500$ samples over $I=200$ iterations. For the novel covariance adaptation strategy, we fix $N_T=50$, which is greater than the dimension $d_x=10$. The results are averaged over $500$ Monte Carlo runs.

Results for this example are shown in Figure \ref{fig: example_1_stuff}. We tested four different schemes which continuously adapt the covariance matrix after each iteration of the algorithm, including the basic AIS algorithm from Table \ref{tab:basic_ais}, with proposal adaptation using eq. \eqref{eq: weighted_empirical_cov} and \eqref{eq: weighted_mean}. In the left-hand plot of Figure \ref{fig: example_1_stuff}, we can see that the basic AIS algorithm adapts to a peaky-proposal distribution, resulting in limited diversity in the generation of samples due to weight degeneracy. The N-PMC algorithm manages to keep the adaptation of the covariance matrix more robust. However, due to the effect of the transformed weights on the adaptation of the location parameter, the proposal distribution remains far from the target. The novel covariance adaptation strategies adapt the best, with the strategy utilizing the weight tempering function performing best. 

The center plot shows the sorted eigenvalues of the adapted covariance matrix and compares to that of the target distribution. The basic AIS method results in a covariance matrix whose eigenvalues are practically $0$. The N-PMC method manages to keep the eigenvalues nonzero, but fails to adapt to the covariance matrix of the target. The novel strategy works best with either choice of weight transformation function. 

The right-hand plot shows the evolution of the Kullback-Liebler (KL) distance between the target and proposal as a function of iteration of the algorithm. This plot emphasizes the advantage of utilizing the weight tempering function over the weight clipping function. We can see that with the weight tempering function, after just $40$ iterations, the KL distance reaches a minimum, whereas the weight clipping function does not reach this value until after $170$ iterations.

% Example 2: Multimodal Target with Good initialization
\subsection{Example 2: High-Dimensional Multimodal Target}
\begin{table*}[t]
	\centering
	%\resizebox{1\columnwidth}{!}{
	\begin{tabular}{|c| |c|c|c|c|c|c|c|}
		\hline
		Method       			& $\sigma=0.5$		& $\sigma=1$      	& $\sigma=2$		& $\sigma=3$      	& $\sigma=5$ 		& $\sigma=7$      	& $\sigma=10$     			\\ \hline\hline
		N-PMC [best]        	& 8.8445       		& 8.0085           	& 5.7241           	& 2.3912           	& 1.8289          	& 5.6683      		& 10.1791 			    	\\ \hline
		N-PMC [worst]        	& 16.3521       	& 17.1335           & 14.8301           & 12.7081           & 9.6968          	& 11.3779      		& 16.3428 			    	\\ \hline
		APIS [best]         	& 2.6422  			& 0.5863          	& 0.3556         	& 3.2842          	& 12.9770          	& 17.9266    		& 22.6413 			    	\\ \hline
		APIS [worst]         	& 3.5907			& 1.1710          	& 0.6811         	& 5.0860         	& 14.3068          	& 18.5037     		& 23.7821 			    	\\ \hline
		DM-PMC [best]   		& 3.0323	       	& 0.9643	        & 1.2454			& 5.7601  			& 14.1088  			& 17.6142  			& 21.8480      				\\ \hline
		DM-PMC [worst]   		& 15.2424       	& 16.7463	        & 18.1355  			& 15.1963 			& 18.6797  			& 23.4776      		& 34.9426 			    	\\ \hline \hline
		\textit{CAIS} [C1] 	& 4.7118   			& 2.8164   			& 1.2644   			& 0.3865   			& 0.4045   			& 1.5499   			& 5.9985 			    	\\ \hline
		\textit{CAIS} [C2] 	& 6.7016  	 		& 2.9347   			& 0.4929   			& 0.1780   			& {\bf 0.3214}   	& {\bf 0.7211}   	& {\bf 4.6918}					\\ \hline
		\textit{CAIS} [T1] 	& 1.7740   			& 0.3834   			& 0.2185   			& 0.2378   			& 0.5632   			& 2.1462   			& 8.8714					\\ \hline
		\textit{CAIS} [T2] 	& {\bf 0.5793}    	& {\bf 0.1931}  	& {\bf 0.0995} 		& {\bf 0.1362}   	& 0.4035   			& 0.9717   			& 7.9077					\\ \hline
	\end{tabular}
	%}
	\caption{Average MSE in approximation of mean of multimodal target distribution. For the novel method (CAIS), we show four different parameter configurations. [C1] uses the clipping function and [T1] uses the  tempering function with $D=25$ and $N_T=0.1\times N$. [C2] uses the clipping function and [T2] uses the  tempering function with $D=50$ and $N_T=0.3\times N$.  }
	\label{tab: example_2_results}
\end{table*}
We now consider the following target distribution
\begin{equation}
\pi(\textbf{x})=\frac{1}{3}\sum_{k=1}^3\mathcal{N}(\textbf{x};\boldsymbol{\nu}_k,\boldsymbol{\Lambda}_k), \quad \textbf{x}\in\mathbb{R}^{10},
\end{equation}
where $\nu_{1,j}=6$, $\nu_{2,j}=-5$ for $j=1,...,10$, and $\boldsymbol{\nu}_3=[1,2,3,4,5,5,4,3,2,1]^\top$. The covariances matrices, $\boldsymbol{\Lambda}_k$ are non-diagonal and were randomly generated from the Wishart distribution. Our objective is to estimate the mean of the target, which is given by, $E_\pi(\textbf{X})=[\frac{2}{3}, 1, \frac{4}{3},\frac{5}{3},2,2,\frac{5}{3},\frac{4}{3},1,\frac{2}{3}]^\top$. The prior mean of each proposal distribution is drawn uniformly, such that $\bmu_{1,d}\sim\mathcal{U}([-10,10]^{10})$. This is considered a good initialization since all modes of the target distribution are contained in this hypercube. We assume a prior covariance matrix for each proposal, $\bSigma_{1,d}=\sigma^2\mathbb{I}_{10}$, where $\mathbb{I}_{10}$ is the $10\times10$ identity matrix. We test for a constant of $M=10000$ samples, $D=\{20,25,40,50,80\}$ mixands for $I=\frac{4\times10^5}{M}$ iterations, where we draw $N=\frac{M}{D}$ samples from each proposal at each iteration. The results are averaged over $500$ Monte Carlo simulations.
The simulation results are summarized in Table \ref{tab: example_2_results}. The novel scheme (CAIS) is compared to three different state-of-the-art AIS algorithms: N-PMC, DM-PMC, and adaptive population importance sampling (APIS). The results indicate that the novel scheme outperforms all three schemes under their best parameter configurations for each value of $\sigma$. 

% Conclusion 
\section{Conclusions}
\label{sec:conclusion}
In this work, we addressed the problem of proposal parameter adaptation in adaptive importance sampling methods for high-dimensional scenarios. The new proposed methods adapt location (mean) parameters using the standard importance weights, while the adaptation of scale (covariance matrix) parameters are conditioned on the effective sample size and utilize transformed importance weights. This robust adaptation improves the performance of AIS and also allows us for better reconstruction of the target distribution as a mixture of kernels with different adapted covariances. Simulation results show excellent performance of the novel strategies compared to other state-of-the-art algorithms. Furthermore, the novel adaptation methods have the potential to extend to other adaptive importance sampling methods, improving their performance for higher-dimensional systems. 

\newpage
\balance 
\bibliographystyle{IEEEbib}
\bibliography{spl_citations}

\end{document}